\documentclass[twoside,twocolumn,9pt]{article}
\usepackage{extsizes}
\usepackage[super,sort&compress,comma]{natbib} 
\usepackage[version=3]{mhchem}
\usepackage[left=1.5cm, right=1.5cm, top=1.785cm, bottom=2.0cm]{geometry}
\usepackage{balance}
\usepackage{mathptmx}
\usepackage{sectsty}
\usepackage{graphicx} 
\usepackage{lastpage}
\usepackage[format=plain,justification=justified,singlelinecheck=false,font={stretch=1.125,small,sf},labelfont=bf,labelsep=space]{caption}
\usepackage{float}
\usepackage{fancyhdr}
\usepackage{fnpos}
\usepackage[english]{babel}
\addto{\captionsenglish}{%
  
}
\usepackage{array}
\usepackage{droidsans}
\usepackage{charter}
\usepackage[T1]{fontenc}
\usepackage[usenames,dvipsnames]{xcolor}
\usepackage{setspace}
\usepackage[compact]{titlesec}
\usepackage{hyperref}
\usepackage{amsmath}        
\usepackage{amssymb}
\usepackage{bm}

\usepackage{epstopdf}

\definecolor{cream}{RGB}{222,217,201}

\begin{document}

\pagestyle{fancy}
\thispagestyle{plain}
\fancypagestyle{plain}{
\renewcommand{\headrulewidth}{0pt}
}

\makeFNbottom
\makeatletter
\renewcommand\LARGE{\@setfontsize\LARGE{15pt}{17}}
\renewcommand\Large{\@setfontsize\Large{12pt}{14}}
\renewcommand\large{\@setfontsize\large{10pt}{12}}
\renewcommand\footnotesize{\@setfontsize\footnotesize{7pt}{10}}
\makeatother

\renewcommand{\thefootnote}{\fnsymbol{footnote}}
\renewcommand\footnoterule{\vspace*{1pt}%
\color{cream}\hrule width 3.5in height 0.4pt \color{black}\vspace*{5pt}} 
\setcounter{secnumdepth}{5}

\makeatletter 
\renewcommand\@biblabel[1]{#1}            
\renewcommand\@makefntext[1]%
{\noindent\makebox[0pt][r]{\@thefnmark\,}#1}
\makeatother 
\renewcommand{\figurename}{\small{Fig.}~}
\sectionfont{\sffamily\Large}
\subsectionfont{\normalsize}
\subsubsectionfont{\bf}
\setstretch{1.125} 
\setlength{\skip\footins}{0.8cm}
\setlength{\footnotesep}{0.25cm}
\setlength{\jot}{10pt}
\titlespacing*{\section}{0pt}{4pt}{4pt}
\titlespacing*{\subsection}{0pt}{15pt}{1pt}

\fancyfoot{}
\fancyfoot[LO,RE]{\vspace{-7.1pt}\includegraphics[height=9pt]{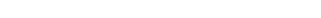}}
\fancyfoot[CO]{\vspace{-7.1pt}\hspace{13.2cm}\includegraphics{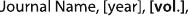}}
\fancyfoot[CE]{\vspace{-7.2pt}\hspace{-14.2cm}\includegraphics{head_foot/RF}}
\fancyfoot[RO]{\footnotesize{\sffamily{1--\pageref{LastPage} ~\textbar  \hspace{2pt}\thepage}}}
\fancyfoot[LE]{\footnotesize{\sffamily{\thepage~\textbar\hspace{3.45cm} 1--\pageref{LastPage}}}}
\fancyhead{}
\renewcommand{\headrulewidth}{0pt} 
\renewcommand{\footrulewidth}{0pt}
\setlength{\arrayrulewidth}{1pt}
\setlength{\columnsep}{6.5mm}
\setlength\bibsep{1pt}

\makeatletter 
\newlength{\figrulesep} 
\setlength{\figrulesep}{0.5\textfloatsep} 

\newcommand{\topfigrule}{\vspace*{-1pt}%
\noindent{\color{cream}\rule[-\figrulesep]{\columnwidth}{1.5pt}} }

\newcommand{\botfigrule}{\vspace*{-2pt}%
\noindent{\color{cream}\rule[\figrulesep]{\columnwidth}{1.5pt}} }

\newcommand{\dblfigrule}{\vspace*{-1pt}%
\noindent{\color{cream}\rule[-\figrulesep]{\textwidth}{1.5pt}} }

\makeatother

\twocolumn[
  \begin{@twocolumnfalse}
\vspace{1em}
\sffamily
\begin{tabular}{m{4.5cm} p{13.5cm} }

\includegraphics{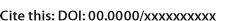} & \noindent\LARGE{\textbf{Accelerating Structure Prediction of Molecular Crystals using Actively Trained Moment Tensor Potential$^\dag$}} \\
\vspace{0.3cm} & \vspace{0.3cm} \\

 & \noindent\large{Nikita Rybin,$^{\ast}$\textit{$^{a, b}$} Ivan Novikov,\textit{$^{a, c,d}$} and Alexander Shapeev\textit{$^{a, b}$}} \\

\includegraphics{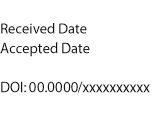} & \noindent\normalsize{Inspired by the recent success of machine-learned interatomic potentials for crystal structure prediction of the inorganic crystals, we present a methodology that exploits Moment Tensor Potentials and active learning (based on maxvol algorithm) to accelerate structure prediction of molecular crystals. Benzene and glycine are used as test systems. Interestingly, among obtained low energy structures of benzene we have found a peculiar polymeric benzene structure.} \\

\end{tabular}

 \end{@twocolumnfalse} \vspace{0.6cm}

  ]

\renewcommand*\rmdefault{bch}\normalfont\upshape
\rmfamily
\section*{}
\vspace{-1cm}


\footnotetext{\textit{$^{a}$~Skolkovo Institute of Science and Technology, Bolshoi bulvar 30, build.1, 121205, Moscow, Russian Federation; $^{\ast}$ E-mail: N.Rybin@skoltech.ru }}
\footnotetext{\textit{$^{b}$~Digital Materials LLC, Kashirskoe rd., build.3/12, 115230, Moscow, Russian Federation}}
\footnotetext{\textit{$^{c}$~Moscow Institute of Physics and Technology, 9 Institutskiy per., Dolgoprudny, Moscow Region, 141701, Russian Federation}}
\footnotetext{\textit{$^{d}$~Emanuel Institute of Biochemical Physics, Kosigina st. 4, 119334 Moscow, Russian Federation}}

\footnotetext{\dag~Supplementary Information available: [details of any supplementary information available should be included here]. See DOI: 00.0000/00000000.}



\section{Introduction}

Molecular crystals exhibit a wide range of applications, including pharmaceuticals, organic electronics, nonlinear optics, and hydrogen storage~\cite{bernstein2020polymorphism, blagden2007crystal, karki2009improving}. Such diverse applicability is a consequence of molecular crystal polymorphism -— the ability of molecules to adopt multiple solid forms. The search for compounds with targeted properties is not only a challenge in inorganic materials science~\cite{zunger2018inverse}, but also in the field of molecular crystals. The experimental trial-and-error approach in the development of compounds with suitable physicochemical properties is resource-consuming, even with the advent of automated chemical laboratories~\cite{Szymanski2023, Tom2024}. 

Computationally guided (\textit{in silico}) crystal structure prediction may become an alternative for exploring the materials' chemical space~\cite{datta2004crystal, beran2023frontiers}. Compared to traditional trial-and-error approaches, which usually depend heavily on the expertise of scientists, computational materials discovery has the advantage of efficiently searching the vast chemical space. The computational modeling of polymorphism has seen many advances in recent years, such as in the generation of reasonable structures~\cite{karamertzanis2007ab}, accurate ranking of these polymorphs~\cite{reilly2016report, neumann2008major, marom2013many}, and even predicting crystal form stability under real-world conditions~\cite{Firaha2023}. Typically, \textit{in silico} crystal structure prediction (for both organic and inorganic crystals) is based on generating a set of structures, followed by geometry optimization (cell shape, volume, and atomic positions) to local minima and choosing the most energetically favorable structures. Traditionally, structural optimization is based on either \textit{ab initio} calculations within the framework of density-functional theory (DFT) or using specifically parameterized force fields. The former approach considerably increases computational resource consumption while allowing for relatively accurate polymorph ranking. The latter approach enables computationally inexpensive exploration of chemical space. However, energy ranking at the classical force field level leaves much to be desired. Machine-learned interatomic potentials bridge this gap and can potentially allow for fast yet accurate screening, as demonstrated for some molecular crystals~\cite{musil2018machine, butler2024machine}.

Inspired by successful machine-learning-based strategies employed for the structure prediction of elemental boron, alloys, and clusters~\cite{PhysRevB.99.064114, Gubaev2019, Wang2022, Manna2023}, in this work we propose to use a Moment Tensor Potential~\cite{Shapeev2016} (MTP) as a model for interatomic interactions to speed up the screening of the chemical space of molecular crystals. The MTP allows for fast and accurate relaxation of molecular crystals, with subsequent lattice energy estimation at a level of accuracy comparable to \textit{ab initio} calculations. The major speedup we obtain is as follows: rather than performing all optimization steps within the DFT framework, we conduct a short prerelaxation of the randomly generated structures with DFT at a fixed unit cell volume and then optimize the structure using MTP until the forces acting on the atoms and stresses vanish. Once the MTP is trained, it is used to perform tight structural optimization without any constraints on cell shape, atomic positions, or cell volume. Consequently, a significant amount of computational resources is saved. In the final stage, we choose the most energetically favorable candidates and perform single-point calculations within the DFT framework using accurate settings.

The construction of MTP is done using active learning approach based on the so-called maxvol algorithm~\cite{Podryabinkin2017}, which allows us to select structures based on a well-established grading system (see Sec.~\ref{al_overview} for details) and construct an accurate machine-learned potential using a minimal amount of data. We conducted extensive experiments to demonstrate MTP's competitive effectiveness and reliability in crystal structure prediction for benzene and glycine molecular crystals. We also evaluated the ability of the MTP trained on a variety of different crystalline structures to predict lattice dynamics properties for the aforementioned molecular crystals. Our work demonstrates the potential for using MTP to make accurate predictions of simple molecular crystals, while we reserve the exploration of more complicated compounds and the evaluation of temperature influence and nuclear quantum effects for future work.

The workflow developed in this work is introduced in Section~\ref{Method}. In particular, subsection~\ref{mtp_overview} contains information on MTP construction, the concept of extrapolation grade is described in subsection~\ref{extrapolation_grade}, and the active learning scheme is detailed in subsection~\ref{al_overview}. Then, in Section~\ref{results}, we benchmark our methodology on ground-state structure prediction for benzene and glycine molecular crystals. In our experience, MTP achieved robustness with a training set comprising roughly a couple of thousand samples, indicating that only a small portion of the actively selected configurations require DFT calculations. Finally, in Section~\ref{discussions}, we discuss limitations and possible extensions of this work. Section~\ref{conclusion_and_outline} sums up our findings.

\section{Methods}\label{Method}

\subsection{Workflow overview}\label{workflow_overview}

\begin{figure}[h!]
	\centering
	\begin{minipage}[h]{1\linewidth}
		\center{\includegraphics[trim={0cm 1cm 14cm 1cm},clip, width=1\linewidth]{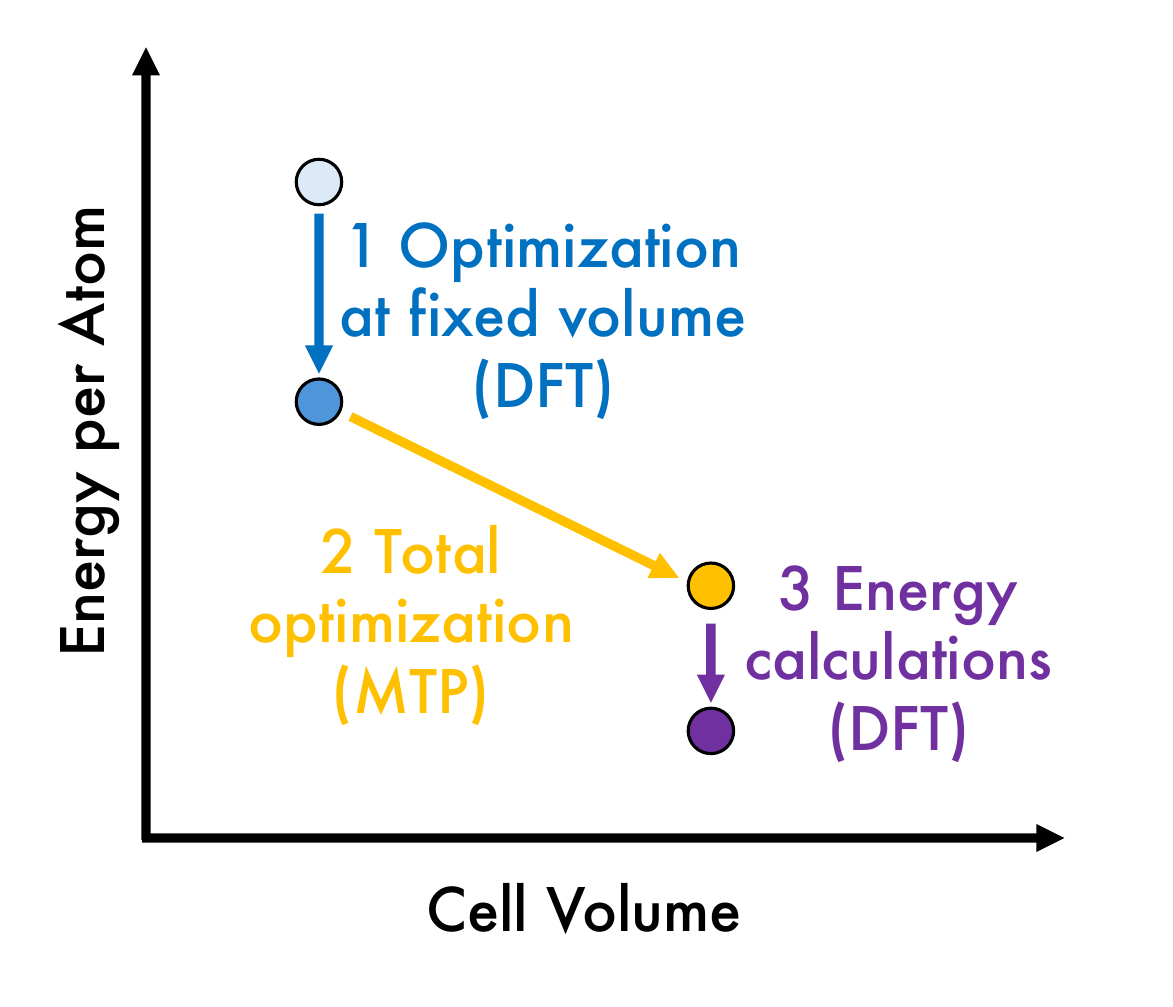}} 
	\end{minipage}
	\caption{Initially, a randomly generated structure is optimized at a fixed volume using DFT. Then, the DFT-optimized structure is fully optimized using MTP to obtain the final structure. At the final stage, DFT calculations with accurate settings for electron density convergence and Brillouin zone sampling grids are used to evaluate the lattice energy. We note that structures optimized with relaxed thresholds for self-consistent field cycle convergence (step 1) are subject to further single-point calculations with tighter settings. Additionally, in step 3, the energy of a structure might also increase.}
	\label{fig:workflow2}
\end{figure}

In this work, we follow the workflow schematically shown in Fig.~\ref{fig:workflow2}. It starts with the generation of a set of initial structures. In principle, one can constrain the search to structures possessing low symmetry and close packing, as preferred by organic crystals~\cite{kitaigorodskii2003organic, zhu2023organic}. However, we decided to generate a set of randomly symmetric structures with a varying number of molecules in unit cells and without imposing constraints on the allowed space groups. This is conveniently done using the PyXtal package~\cite{fredericks2021pyxtal}.

The initial volume of the randomly generated structures can be wisely chosen based on the van der Waals radii of atoms; however, these randomly generated structures have large stresses and forces acting on the cell and atoms, respectively. Hence, we perform optimization of the cell shape and atomic positions while keeping the volume constant. At this step (labeled 1 in Fig.~\ref{fig:workflow2}), the accuracy of the self-consistent field convergence threshold and the density of the k-point mesh used for Brillouin zone sampling are far from converged. Once structures are pre-converged, we perform accurate calculations of energy, forces, and stresses within the scope of DFT. This serves as an initial dataset for MTP training. Then, we start the optimization (relaxation) of the structural parameters without constraints, i.e., atomic positions, cell shape, and cell volume can change during optimization. At this step (labeled 2 in Fig.~\ref{fig:workflow2}) MTP is used as a model for interatomic interactions. 

\begin{figure}[h!]
	\centering
	\begin{minipage}[h]{1\linewidth}
		\center{\includegraphics[trim={0cm 1cm 14cm 0cm},clip, width=1\linewidth]{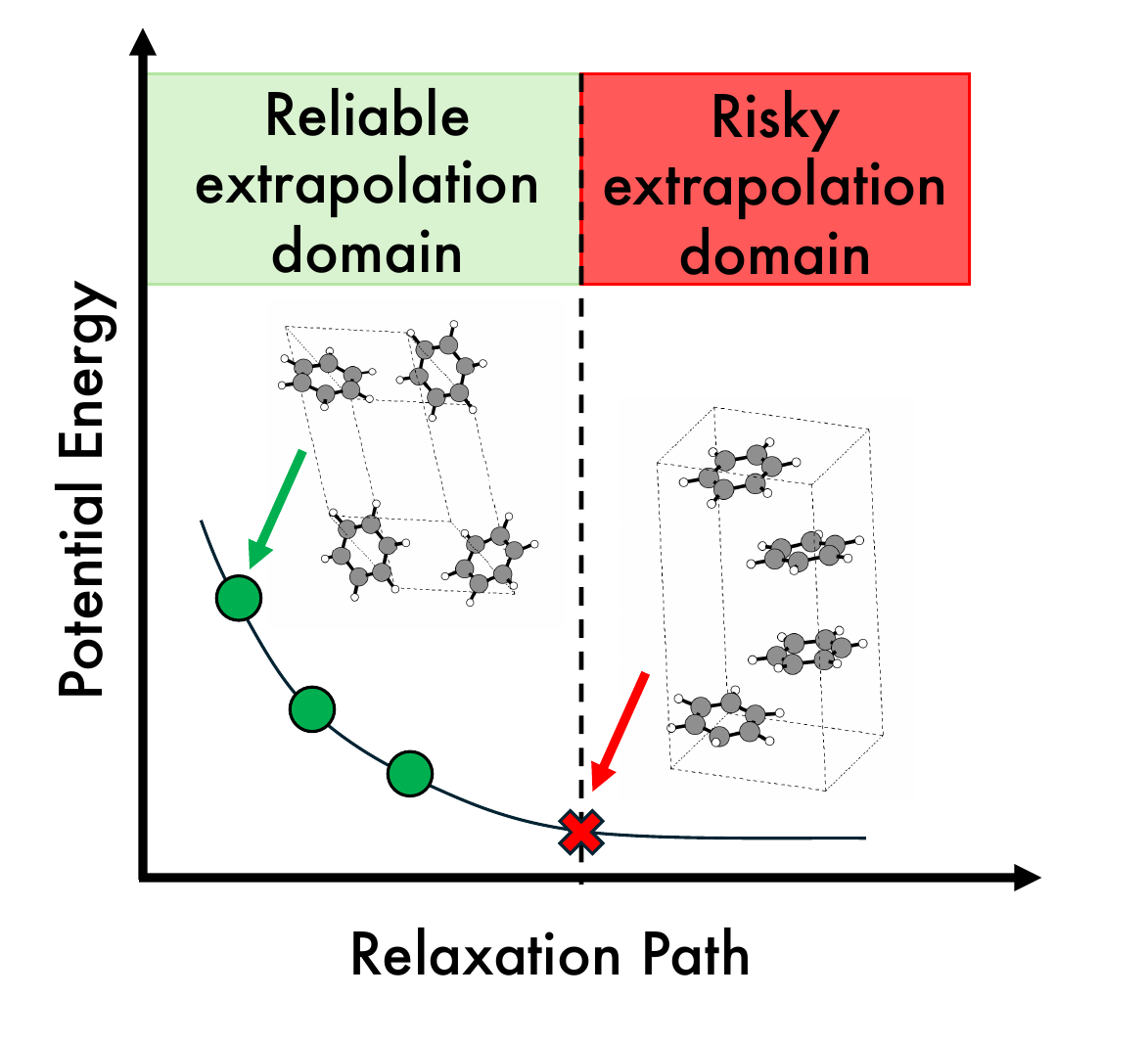}} 
	\end{minipage}
	\caption{Graphical illustration of the structure optimization process during which we perform active learning of MTP. The reliable extrapolation domain is highlighted in green and corresponds to the area of chemical space that can be accurately evaluated using MTP. The risky extrapolation region, highlighted in red, corresponds to the area of chemical space (the potential energy surface, which is shown as a black line) where the potential cannot be used to perform reliable structural optimization.}
	\label{fig:workflow1}
\end{figure}

During MTP-based structural optimization, we monitor that at each step of relaxation, the potential does not extrapolate beyond the predicted energy, forces, and stresses. If the potential heavily extrapolates (the extrapolation grade is evaluated at each relaxation step, as discussed in subsections ~\ref{extrapolation_grade} and ~\ref{al_overview}) on energies, forces, and stresses predictions, it cannot be used for structural optimization purposes. This is schematically shown in Fig.~\ref{fig:workflow1}. Once the potential approaches a risky extrapolation region, we halt the relaxation and add the structures for which the potential extrapolates to the training set. Next, we conduct DFT calculations for the configurations added to the training set and re-fit the potential. We continue this process, i.e., relaxation with extrapolation control, selection of extrapolative configurations, DFT calculations for them, and re-fitting the potential until no extrapolative configurations appear during the relaxation. Details of this workflow are provided below in subsections ~\ref{extrapolation_grade} and ~\ref{al_overview}. The on-the-fly generation of the training set based on the estimation of the extrapolation grades during MD-based configurational space sampling was previously tested for solids~\cite{rybin2024moment} and liquid electrolytes~\cite{rybin2024thermophysical}. On-the-fly learning using structure optimization was also previously used for generating MTPs of alloys~\cite{Gubaev2019}.

As will be shown, the obtained potentials optimize structures to local minima and have mean absolute errors in energy comparable to the differences among energetically favorable polymorphs. For a sanity check, once the MTP-based geometry optimization is completed, we perform accurate DFT-based calculations of the total energy of the most energetically favorable polymorphs. This is step number 3, as shown in Fig.~\ref{fig:workflow2}. Obviously, polymorph ranking can be done at different levels of theory. For example, using the hybrid PBE0 exchange-correlation functional~\cite{Adamo1999} and MBD correction~\cite{PhysRevLett.108.236402} ensures accurate polymorph ranking. This would increase the cost of force calculations for an average-sized system by about 50 \%, but considering the efficiency of our active learning methodology, which will be demonstrated below, this might be affordable. In extreme cases, ranking can be done with sub-kilojoule per mole accuracy using quantum chemistry methods~\cite{Yang2014}. 

\subsection{Moment Tensor Potential}\label{mtp_overview}

In this work, we used Moment Tensor Potential (MTP) implemented in the MLIP-2 package~\cite{Novikov2021} to perform molecular crystal structure optimization. In the scope of MTP, the potential energy of an atomic system is defined as a sum of the energies of atomic environments of the individual atoms:

\[
    E^{\text{MTP}} = \sum_{i=1}^{N} V({\mathfrak{\bm n}}_{i}),
\]
where the index $i$ labels $N$ atoms of the system, ${\mathfrak{\bm n}}_{i}$ denotes the local atomic neighborhood around atom \textit{i} within a certain cut-off radius $R_\text{cut}$, and the function $V$ is the energy of atomic neighborhood: 
\[
    V({\mathfrak{\bm n}}_{i}) = \sum_{\alpha} \xi_{\alpha} B_{\alpha}({\mathfrak{\bm n}}_{i}).
\]
Here $\xi_{\alpha}$ are the linear parameters to be fitted and $B_{\alpha}({\mathfrak{\bm n}}_{i})$ are the basis functions that will be defined below. As fundamental symmetry requirements, all descriptors for atomic environment have to be invariant to translation, rotation, and permutation with respect to the atomic indexing. Moment tensors descriptors $M_{\mu, \nu}$ satisfy this requirements and are used as representations of atomic environments: 
\[
    M_{\mu, \nu}\left({\mathfrak{\bm n}}_{i}\right)=\sum_j f_\mu\left(\left|r_{i j}\right|, z_i, z_j\right) \underbrace{{\bm r}_{i j} \otimes \ldots \otimes \bm{r}_{i j}}_{\nu \text { times }},
\]
where the index $j$ goes through all the neighbors of atom $i$. The symbol ``$\otimes$'' stands for the outer product of vectors, thus ${\bm r}_{i j} \otimes \cdots \otimes {\bm r}_{i j}$ is the tensor of rank $\nu$ encoding the angular part. 
The function $f_\mu$ represents the radial component of the moment tensor:
\[
f_\mu\left(\left|{\bm r}_{i j}\right|, z_i, z_j\right)=\sum_{\beta} c_{\mu, z_i, z_j}^{(\beta)} Q^{(\beta)}(\left|{\bm r}_{i j}\right|),
\]
where $z_i$ and $z_j$ denote the atomic species of atoms $i$ and $j$, respectively, ${\bm r}_{ij}$ describes the positioning of atom $j$ relative to atom $i$, $c_{\mu, z_i, z_j}^{(\beta)}$ are the radial parameters to be fitted and 
\[
Q^{(\beta)}(\left|{\bm r}_{i j}\right|)=T^{(\beta)}(\left|{\bm r}_{i j}\right|)\left(R_{\text {cut }}-\left|{\bm r}_{i j}\right|\right)^2
\]
are the radial functions consisting of the Chebyshev polynomials $T^{(\beta)}(\left|{\bm r}_{i j}\right|)$ on the interval $[R_\text{min},  R_\text{cut}]$ and the term ($R_\text{cut} - \left|{\bm r}_{i j}\right|)^2$ that is introduced to ensure a smooth cut-off to zero. The descriptors $M_{\mu, \nu}$ taking $\nu$ equal to $0, 1, 2, \ldots$ are tensors of different ranks that allow one to define basis functions as all possible contractions of these tensors to a scalar, for instance:
\[
\begin{aligned}
& B_0\left({\mathfrak{\bm n}}_{i}\right)=M_{0,0}\left({\mathfrak{\bm n}}_{i}\right), \\
& B_1\left({\mathfrak{\bm n}}_{i}\right)=M_{0,1}\left({\mathfrak{\bm n}}_{i}\right) \cdot M_{0,1}\left({\mathfrak{\bm n}}_{i}\right), \\
& B_2\left({\mathfrak{\bm n}}_{i}\right)=\left(M_{2,2}\left({\mathfrak{\bm n}}_{i}\right) M_{2,1}\left({\mathfrak{\bm n}}_i\right)\right) \cdot  M_{0,1}\left({\mathfrak{\bm n}}_{i}\right), \\
& \ldots
\end{aligned}
\]

\noindent However, the number of contractions yielding a scalar, i.e., basis functions $B_{\alpha}$ is infinite. In order to restrict the number of basis functions in the MTP functional form, we introduce the level of moment tensor descriptors ${\rm lev}M_{\mu, \nu}$ = 2 + 4$\mu$ + $\nu$. If $B_{\alpha}$ is obtained from $M_{\mu_1, \nu_1}$, $M_{\mu_2, \nu_2}$ ,
$\dots$, then ${\rm lev}B_{\alpha}$ = $(2 + 4\mu_1 + \nu_1)$ + $(2 + 4\mu_2 + \nu_2)$ + $\dots$. By including all basis functions with ${\rm lev}B_{\alpha} \leq d$, we obtain MTP of level $d$ including a finite number of the basis functions $B_{\alpha}$. We denote the total set of parameters to be found by ${\bm \theta} = (\{ \xi_{\alpha} \}, \{ c^{(\beta)}_{\mu, z_i, z_j} \})$ and the MTP energy of a configuration $E^{\text{MTP}} = E({\bm \theta})$.

\subsection{Extrapolation grade of configurations}\label{extrapolation_grade}

For automated constructing a training set for MTP fitting we calculate the so-called extrapolation grade of configurations obtained during the structural relaxation. For estimating the extrapolation grade of configurations we conduct the following steps. Assume we have $K$ configurations in an initial training set with energies $E^{\rm DFT}_k$, forces ${\bm f}^{\rm DFT}_{i,k}$, and stresses $\sigma^{\rm DFT}_{i,k}$, $k = 1, \ldots, K$ calculated within DFT framework. We start with fitting an initial MTP, i.e., with finding the optimal parameters ${\bm{\bar{\theta}}}$ by solving the following minimization problem:
\begin{equation} \notag
\begin{array}{c}
\displaystyle
\sum \limits_{k=1}^K \Bigl[ w_{\rm e} \left(E^{\rm DFT}_k - E_k({\bm {\theta}}) \right)^2 + w_{\rm f} \sum_{i=1}^N \left| {\bm f}^{\rm DFT}_{i,k} - {\bm f}_{i,k}({\bm {\theta}}) \right|^2 
\\
\displaystyle
+ w_{\rm s} \sum_{i=1}^6 \left| \sigma^{\rm DFT}_{i,k} - \sigma_{i,k}({\bm {\theta}}) \right|^2 \Bigr] \to \operatorname{min},
\end{array}
\end{equation}
where $w_{\rm e}$, $w_{\rm f}$, and $w_{\rm s}$ are non-negative weights expressing the importance of energies, forces, and stresses in the above minimization problem. 

After finding the optimal parameters ${\bm{\bar{\theta}}}$ we compose the following matrix
\[
\mathsf{B}=\left(\begin{matrix}
\frac{\partial E_1\left( {\bm{\bar{\theta}}} \right)}{\partial \theta_1} & \ldots & \frac{\partial E_1\left( {\bm{\bar{\theta}}} \right)}{\partial \theta_m} \\
\vdots & \ddots & \vdots \\
\frac{\partial E_K\left( {\bm{\bar{\theta}}} \right)}{\partial \theta_1} & \ldots & \frac{\partial E_K\left( {\bm{\bar{\theta}}} \right)}{\partial \theta_m} \\
\end{matrix}\right),
\]
where each row corresponds to a particular configuration. Next, we construct a subset of configurations yielding the most linearly independent rows (physically it means geometrically different configurations) in $\mathsf{B}$. This is equivalent to finding a square $m \times m$ submatrix $\mathsf{A}$ of the matrix $\mathsf{B}$ of maximum volume (maximal value of $|{\rm det(\mathsf{A})}|$). To achieve this, we use the so-called maxvol algorithm \cite{zamarashkin2010-maxvol}. The resulting matrix $\mathsf{A}$ is called an active set. To determine whether a given configuration $\bm x^*$ obtained during structural relaxation is representative, we calculate the extrapolation grade $\gamma(\bm x^*)$ defined as
\begin{equation} \label{Grade}
\begin{array}{c}
\displaystyle
\gamma(\bm x^*) = \max_{1 \leq j \leq m} (|c_j|), ~\rm{where}
\\
\displaystyle
c = \left( \dfrac{\partial E}{\partial \theta_1} (\bm \theta, \bm x^*) \ldots \dfrac{\partial E}{\partial \theta_m} (\bm \theta, \bm x^*) \right) \mathsf{A}^{-1}.
\end{array}
\end{equation}
This grade defines the maximal factor by which the determinant $|{\rm det(\mathsf{A})}|$ can increase if ${\bm x^*}$ is added to the training set. Thus, if the configuration $\bm x^*$ is a candidate for adding to the training set then $\gamma(\bm x^*) \geq \gamma_{\rm th}$, where $\gamma_{\rm th} \geq 1$ is an adjustable threshold parameter which controls the value of permissible extrapolation. Otherwise, the configuration is not representative. In this work we used $\gamma_{\rm th} = 2.1$ and once extrapolation grade exceeds 10 the relaxation was stopped. Such values were chosen based on the previous benchmarks~\cite{Podryabinkin2017,Novikov2021}.

\subsection{Active learning scheme}\label{al_overview}

Here, we describe an active learning scheme allows us to automatically construct a training set and fit MTP for predicting molecular crystals. The active learning workflow is schematically demonstrated in Fig.~\ref{fig:workflow1}. As it was mentioned above, at each step of the structural relaxation the algorithm assesses the extrapolation grade $\gamma$ of the atomic configuration based solely on atomic coordinates. Configurations with $\gamma < 2.1$ are considered to be accurately calculated, while the ones with $2.1 \leq \gamma \leq 10$ are added to the so-called preselected set (green circles in Fig.~\ref{fig:workflow1}, the reliable extrapolation region). When $\gamma$ exceeds 10 (red cross in Fig.~\ref{fig:workflow1}), the structural optimization halts because we have approached the risky extrapolation region. After that, we calculate the matrix with the derivatives of energies for the preselected configurations with respect to MTP parameters and combine it with the matrix $\mathsf{A}$ (active set). Using the maxvol algorithm, we select the rows corresponding to different and representative configurations in the combined matrix, calculate the new active set, and incorporate all the novel different and representative configurations into the training set. Next, we conduct DFT calculations and obtain DFT energies, forces, and stresses for the novel configurations. Finally, we re-fit the potential. This procedure repeats until all structural optimizations can run without preselected configurations, i.e., without extrapolative configurations. This active learning scheme plays a crucial role in enhancing the robustness of any MLIP, and MTP in particular. Its employment not only ensures the applicability of MTP to structural optimization, but also significantly reduces the number of required DFT calculations.  


\subsection{Computational details}

The MTP utilized in this study is trained using data computed within the DFT framework. All DFT calculations were carried out using Vienna \textit{ab initio} simulations package (VASP)~\cite{Kresse1996} with the projector augmented wave method~\cite{Kresse1999}. The Perdew-Burke-Ernzerhof generalized gradient approximation (PBE-GGA)~\cite{Perdew1996} was employed for the exchange–correlation functional, and the D3 method~\cite{Grimme2010} was utilized to account for dispersion correction. For structural pre-optimization, we used a plane-wave energy cut-off of 520~eV, while accurate energy evaluation was done with a cut-off of 600~eV. The initial structural optimization was carried out at fixed volumes, and Brillouin zone (BZ) was sampled only in the $\Gamma$-point. Later, the lowest energy structures determined by MTP were then subjected to further optimization using a stringent set of parameters, which we use for precise energy evaluation. For accurate total energy evaluation we used $\Gamma$-centered k-meshes with a resolution of 2$\pi$ $\times$ 0.05 \AA$^{-1}$ for BZ sampling, ensuring the convergence of a total energy of 1~meV/atom.

In this work, we used MTP of the 20-th level, i.e., with 288 basis functions. We took a cut-off radius of 5 \AA ~in order to ensure a non-zero interaction between each pair of atoms in a single molecule of benzene and glycine.

\section{Results}\label{results}

Here, we discuss benchmarks of our methodology on systems with well-known ground-states structures -- benzene and glycine. Benzene (C$_{6}$H$_{6}$) is the simplest aromatic compound and it has a purely planar molecule, the packing of which is stabilized by $\pi$--$\pi$ interactions. The crystal structure of benzene is one of the most basic and most actively investigated structures~\cite{ciabini2005high, ciabini2007triggering, wen2011benzene, raiteri2005exploring}. Glycine, with the formula C$_{2}$H$_{5}$NO$_{2}$, is the smallest of twenty aminoacids commonly found in proteins~\cite{finkelstein2016protein}. The polymorphism of glycine was intensely studied~\cite{zhu2012constrained, chisholm2005ab, hamad2008clustering, dawson2005effect, moggach2008high, boldyreva2003structural, perlovich2001polymorphism}. Glycine is known to crystallize in four polymorphs with space groups $P2_{1}/c$, $P2_{1}$, $P3_{2}$ and $P2_{1}/c$, which are labeled $\alpha$, $\beta$, $\gamma$, and $\sigma$, respectively~\cite{chisholm2005ab}. The $\alpha$, $\beta$, $\gamma$ phases are found at ambient pressure (while $\sigma$ only under compression~\cite{dawson2005effect}), with $\alpha$, and $\beta$ phases being metastable with respect to the $\gamma$ phase. 

For both test cases, we generated datasets consisting of 3000 candidate structures using the random-symmetric molecular crystal structure generator. During this structure generation, crystal structures comprised from two to six molecules per unit cell. The angles and cell dimensions of each structure were not restricted. Additionally, minimum separations between atoms were specified based on the van der Waals radii of ions (see section ~\ref{Method} for details). In the case of glycine we used zwitterionic form of a molecule during structure generation. All these structures were prerelaxed using light settings for self-consistent field (SCF) cycle convergence to minimize exceptionally high stresses and forces acting on unit cell and atoms, respectively.

The potential fitting and active learning procedure are identical in the benzene and glycine cases. We randomly selected 200 structures with $Z=2-4$ molecules to fit the initial MTP. The task was then to tightly relax all compounds that were left in the datasets, while increasing the training set size -- following the active learning methodology presented in subsection ~\ref{al_overview}. In the case of benzene, during the active learning procedure, the algorithm selected 1352 structures, and the resulting potential has 3.47~meV/atom energy mean absolute errors (MAE) and 43.4 meV/\AA ~force MAE. In the case of glycine, the algorithm selected 3127 samples, and the resulting potential has 3.09 meV/atom energy MAE and 40.9~meV/\AA ~force MAE. We should note that in the case of benzene, it was possible to achieve energy MAE less than 5 meV/atom using the MTP with 450 parameters, while in the case of glycine, it was achieved with the MTP consisting of 932 parameters. Information about the potential, training sets, and errors is shown in Table~\ref{tab:test_train_errors}. Based on the MTP ranking, we selected compounds with energies higher by 50 meV/atom than the ground state compound at most. For these structures, we performed single-point DFT calculations with accurate settings (see Section ~\ref{Method} for computational settings).

\begin{table}[h]
 \caption{Number of parameters in the used MTP, number of samples, and mean absolute errors on energies and forces after additional structures were added.} 
 \centering
\begin{tabular}{c|c|c}
& benzene & glycine     \\ \hline
\begin{tabular}[c]{@{}c@{}}\# Parameters\\ in MTP\end{tabular}                  & 450     & 932         \\ \hline
\begin{tabular}[c]{@{}c@{}}\# Samples \\ in the \\ training set\end{tabular}    & 1352  & 3127 \\ \hline
\begin{tabular}[c]{@{}c@{}}Train MAE \\ on Energy \\ (meV/atom)\end{tabular}    & 3.47    & 3.09 \\ \hline
\begin{tabular}[c]{@{}c@{}}Train MAE \\ on Forces\\ (meV/Angstrom)\end{tabular} & 43.4    & 40.9 \\ \hline
\end{tabular}
\label{tab:test_train_errors}
\end{table}

\begin{figure*}[h!]
	\centering
 	\begin{minipage}[h]{1\linewidth}
		\center{\includegraphics[trim={0cm 2cm 0cm 0cm},clip, width=1\linewidth]{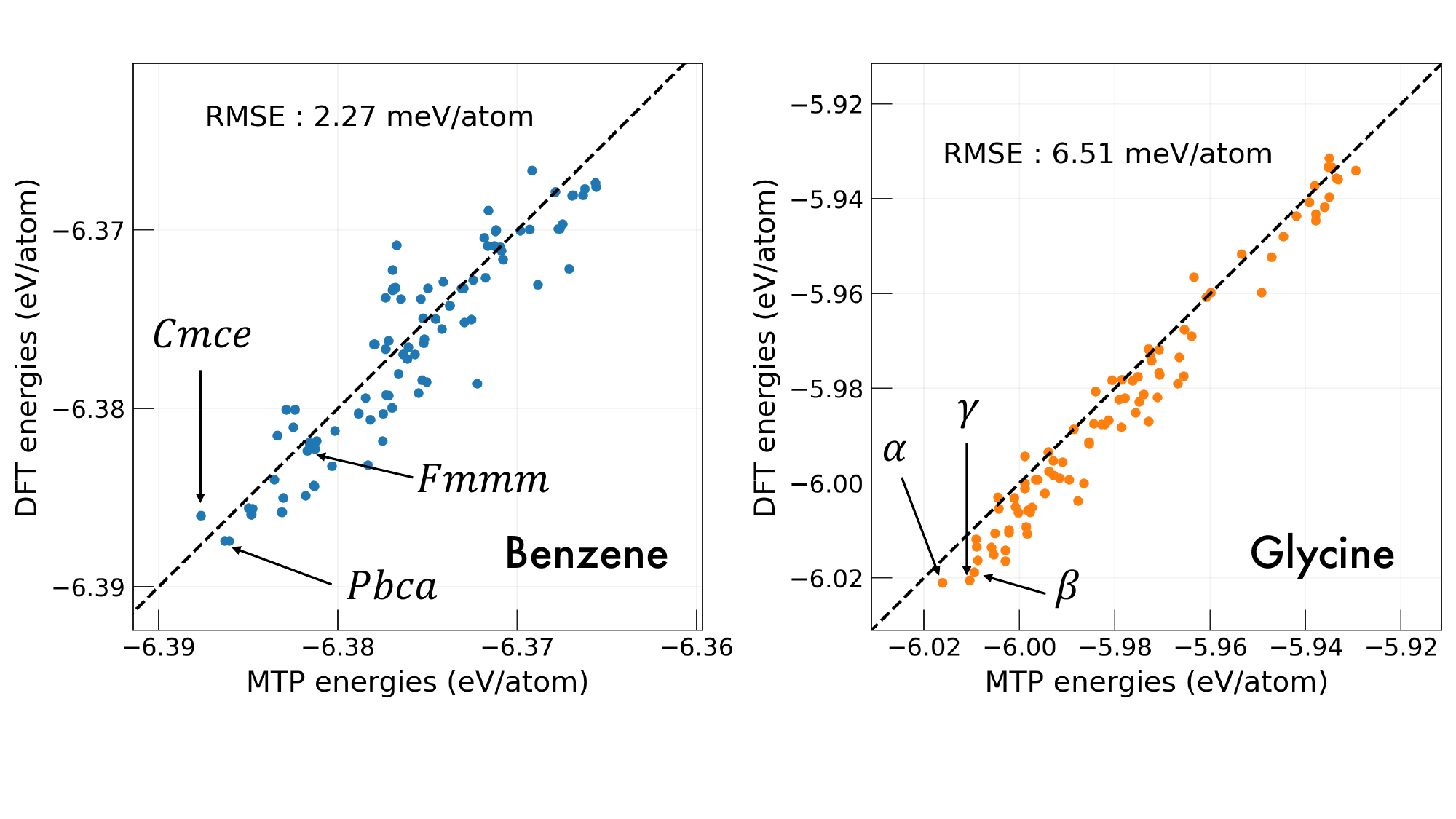}}
	\end{minipage}
	\caption{Comparison between DFT-calculated and MTP-calculated energies of the most energetically favorable benzene (on the left) and glycine (on the right) polymorphs. }
	\label{fig:benzene_glycine_dft_vs_mtp}
\end{figure*}

In general, molecular crystals exhibit complex polymorphic energy landscapes with numerous structures located within an energy window of a few kilojoules per mole~\cite{reilly2016report}. Hence, a useful model should have accuracy comparable with DFT to accurately rank polymorphs. As it is demonstrated in Fig.~\ref{fig:benzene_glycine_dft_vs_mtp}, energies obtained by MTP for MTP-optimized structures are in good agreement with the energies obtained using accurate DFT calculations. The root-mean-square errors are 2.27~meV/atom and 6.51~meV/atom in the case of benzene and glycine, respectively. 

\begin{figure}[h!]
	\centering
	\begin{minipage}[h]{0.95\linewidth}
		\center{\includegraphics[trim={0cm 6cm 18cm 0cm},clip, width=1\linewidth]{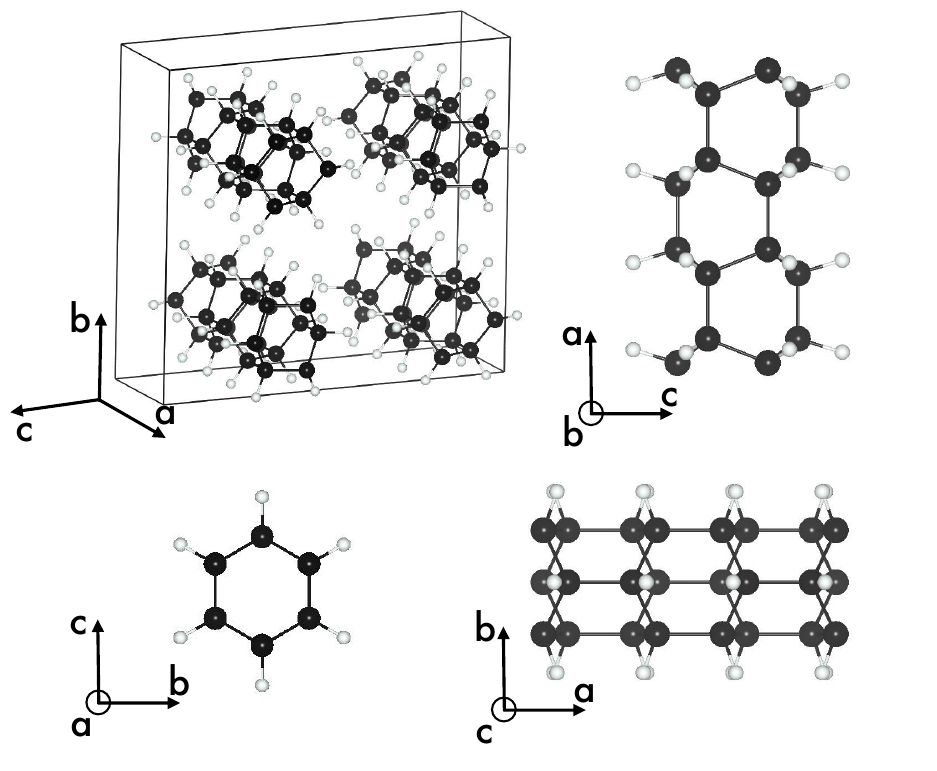}} 
	\end{minipage}
	\caption{Structure of polymerized benzene with the $Fmmm$ space group.}
	\label{fig:benzene_fmmm}
\end{figure}

In the case of benzene, we found out that the ground state structure is the orthorhombic structure with $Pbca$ space group in the case of DFT-based ranking and the structure with $Cmce$ space group in the case of MTP-based ranking. DFT-based ranking is in good agreement with the literature~\cite{ciabini2005high, ciabini2007triggering, wen2011benzene, raiteri2005exploring} (we also checked the structural parameters). The structure with $Cmce$ space group appeared previously in the literature~\cite{raiteri2005exploring} and it was even suggested that there is a second-order phase transition between $Cmce$-benzene and $Pbca$-benzene~\cite{10.1063/1.454809}. Hence, we believe that the inability of MTP to distinguish these polymorphs is acceptable since it was also trained on the dataset of different configurations. Its curious that among energetically favorable benzene structure we have a number of structures, which appear on the benzene phase diagram~\cite{wen2011benzene}, e.g., structures with $P2_{1}/c$, $P2_{1}$ and $P4_{3}2_{1}2$ space groups, and our algorithm has found a polymeric benzene structure with $Fmmm$ space group, shown in Fig.~\ref{fig:benzene_fmmm}. Polymerization of benzene at high pressures was theoretically predicted and described before~\cite{wen2011benzene}, but the polymers predicted previously are structurally different from the one obtained in this study. Our potential was calibrated at normal conditions and may fail at high pressure. Thus, we leave phase-diagram exploration of benzene for future work.

In the case of glycine, all three most energetically favorable structures, i.e., $\alpha$-, $\beta$-, and $\gamma$-glycine, were found as the most energetically favourable. This highlights the power of our search methodology. However, both MTP and DFT-based rankings show that $\alpha$-glycine possesses the lowest enthalpy, while the $\beta$ and $\gamma$ phases are slightly higher in energy. Yet the experimental results demonstrated the relative thermodynamic stability to be $\gamma$ > $\alpha$ > $\beta$. This is not an unexpected result and was previously observed in other DFT-based studies of glycine molecular crystal~\cite{zhu2012constrained}, which points to the need for better ways of computing intermolecular interaction energies (see section ~\ref{discussions} for a more detailed overview of limitations).


We also checked the ability of MTP potential to compute dynamical stability of the lowest lying polymorphs. According to our calculations  $\alpha$-, $\beta$-, and $\gamma$-glycine phases are dynamically stable. Moreover, next two phases are also dynamically stable, since no imaginary frequencies are observed in their phonon spectrum as shown in Fig.S1 in the Supplementary Information. 

\section{Discussion}\label{discussions}

In this work, we did not focus on the hyperparameter optimization of our model. Although it might be beneficial for reducing training errors, we believe that the true yield can be obtained by using more advanced architectures of the ML potentials. For example, the Equivariant Tensor Network (ETN) potential, recently developed by one of us, allows for the usage of even smaller amounts of data to obtain similar training errors~\cite{hodapp2023equivariant}. Thus, we would rather transfer to the novel architecture in future work than explore hyperparameter tuning of MTP.

There are several limitations that we would like to discuss. In particular, while the results for benzene and glycine are promising and pave the way for applying MTP to accelerate molecular crystals structure prediction, these systems are relatively simple. In future work, we plan to try our methodology on more challenging systems.

From our perspective, one of the major challenges ahead is associated with the explicit treatment of long-range interactions in machine-learned interatomic potentials. Although some work is already done in this direction, we believe that this problem is far from being solved~\cite{Ko2021, Shaidu2024, Grisafi2019}. In future work, we plan to explicitly implement long-range interactions in the functional form of our MLIPs.

Last but not least, all the calculations described here were performed at zero temperature. Hence, lattice thermal expansion and entropy contributions were neglected. In principle, once a list of the most interesting polymorphs is formed, one can create a machine-learned potential that can be used for obtaining Gibbs free energy from molecular dynamics (MD) simulations. A potential trained in this manner should demonstrate the ability to robustly conduct MD simulations for at least hundreds of picoseconds. The ability of the MTP to obtain the phonon spectrum of molecular crystals presented here allows us to conclude that the MTP approach can properly describe the lattice dynamics of these systems. As we have recently shown, the active learning procedure based on the maxvol algorithm perfectly suits such a task~\cite{rybin2024thermophysical, rybin2024moment}. Recently, it was demonstrated how neural network-based MLIP can be used for such a task to catch different Free Energy terms (lattice thermal expansion, nuclear quantum effect, entropy term)~\cite{kapil2022complete} in simple molecular crystals (also benzene and glycine were used as examples). In future work, we will check if our active learning scheme can be used to generate efficient machine-learned models and if it could be used in path integral MD, especially when coupled with sophisticated enhanced sampling techniques. Our confidence in combining PIMD and active learning of MTP stems from the fact that we have already successfully combined another method of quantum molecular dynamics, namely, ring-polymer molecular dynamics, and active learning of MTP and used this combination for predicting chemical reaction rates at different temperatures \cite{novikov2024-rpmd-oh-hbr}.


\section{Conclusions}\label{conclusion_and_outline}

We present a workflow for accelerated molecular crystal structure prediction obtained by utilizing Moment Tensor Potential as a model for interatomic interactions. A key element of our workflow is the active-learning methodology, which allows us to generate accurate potentials with minimal data utilization. We demonstrated the effectiveness of our approach by predicting ground-state polymorphs of benzene and glycine. Interestingly, among the obtained low-energy structures of benzene, we found a peculiar polymeric benzene structure. Since polymerization of benzene was previously found under pressure, we highlight that our computational screening samples the chemical space meticulously. In addition, we have shown that the potential trained on the different polymorphs could be used for lattice dynamics calculations of the energetically favorable polymorphs. Consequently, this potential can be later used as a starting point for lattice dynamics simulations.

We note that beyond the very important application to molecular crystals, the presented workflow could also become relevant in other domains of structure prediction, such as inorganic solids or nanoclusters, and paves the way for \textit{in silico} prediction of materials with tailored properties.

\section{Author contributions}
N.R. initiated the study and performed crystal structure prediction calculations. I.N. performed active learning of MTP. A.S. supervised the project. All authors participated in the manuscript writing, reviewing, and editing.


%
%



\balance


\clearpage
\bibliography{rsc} 

\providecommand*{\mcitethebibliography}{\thebibliography}
\csname @ifundefined\endcsname{endmcitethebibliography}
{\let\endmcitethebibliography\endthebibliography}{}
\begin{mcitethebibliography}{54}
\providecommand*{\natexlab}[1]{#1}
\providecommand*{\mciteSetBstSublistMode}[1]{}
\providecommand*{\mciteSetBstMaxWidthForm}[2]{}
\providecommand*{\mciteBstWouldAddEndPuncttrue}
  {\def\EndOfBibitem{\unskip.}}
\providecommand*{\mciteBstWouldAddEndPunctfalse}
  {\let\EndOfBibitem\relax}
\providecommand*{\mciteSetBstMidEndSepPunct}[3]{}
\providecommand*{\mciteSetBstSublistLabelBeginEnd}[3]{}
\providecommand*{\EndOfBibitem}{}
\mciteSetBstSublistMode{f}
\mciteSetBstMaxWidthForm{subitem}
{(\emph{\alph{mcitesubitemcount}})}
\mciteSetBstSublistLabelBeginEnd{\mcitemaxwidthsubitemform\space}
{\relax}{\relax}

\bibitem[Bernstein(2020)]{bernstein2020polymorphism}
J.~Bernstein, \emph{Polymorphism in molecular crystals 2e}, International Union of Crystal, 2020, vol.~30\relax
\mciteBstWouldAddEndPuncttrue
\mciteSetBstMidEndSepPunct{\mcitedefaultmidpunct}
{\mcitedefaultendpunct}{\mcitedefaultseppunct}\relax
\EndOfBibitem
\bibitem[Blagden \emph{et~al.}(2007)Blagden, de~Matas, Gavan, and York]{blagden2007crystal}
N.~Blagden, M.~de~Matas, P.~T. Gavan and P.~York, \emph{Advanced drug delivery reviews}, 2007, \textbf{59}, 617--630\relax
\mciteBstWouldAddEndPuncttrue
\mciteSetBstMidEndSepPunct{\mcitedefaultmidpunct}
{\mcitedefaultendpunct}{\mcitedefaultseppunct}\relax
\EndOfBibitem
\bibitem[Karki \emph{et~al.}(2009)Karki, Fri{\v{s}}{\v{c}}i{\'c}, F{\'a}bi{\'a}n, Laity, Day, and Jones]{karki2009improving}
S.~Karki, T.~Fri{\v{s}}{\v{c}}i{\'c}, L.~F{\'a}bi{\'a}n, P.~R. Laity, G.~M. Day and W.~Jones, \emph{Advanced materials}, 2009, \textbf{21}, 3905--3909\relax
\mciteBstWouldAddEndPuncttrue
\mciteSetBstMidEndSepPunct{\mcitedefaultmidpunct}
{\mcitedefaultendpunct}{\mcitedefaultseppunct}\relax
\EndOfBibitem
\bibitem[Zunger(2018)]{zunger2018inverse}
A.~Zunger, \emph{Nature Reviews Chemistry}, 2018, \textbf{2}, 0121\relax
\mciteBstWouldAddEndPuncttrue
\mciteSetBstMidEndSepPunct{\mcitedefaultmidpunct}
{\mcitedefaultendpunct}{\mcitedefaultseppunct}\relax
\EndOfBibitem
\bibitem[Szymanski \emph{et~al.}(2023)Szymanski, Rendy, Fei, Kumar, He, Milsted, McDermott, Gallant, Cubuk, Merchant, Kim, Jain, Bartel, Persson, Zeng, and Ceder]{Szymanski2023}
N.~J. Szymanski, B.~Rendy, Y.~Fei, R.~E. Kumar, T.~He, D.~Milsted, M.~J. McDermott, M.~Gallant, E.~D. Cubuk, A.~Merchant, H.~Kim, A.~Jain, C.~J. Bartel, K.~Persson, Y.~Zeng and G.~Ceder, \emph{Nature}, 2023, \textbf{624}, 86--91\relax
\mciteBstWouldAddEndPuncttrue
\mciteSetBstMidEndSepPunct{\mcitedefaultmidpunct}
{\mcitedefaultendpunct}{\mcitedefaultseppunct}\relax
\EndOfBibitem
\bibitem[Tom \emph{et~al.}(2024)Tom, Schmid, Baird, Cao, Darvish, Hao, Lo, Pablo-García, Rajaonson, Skreta, Yoshikawa, Corapi, Akkoc, Strieth-Kalthoff, Seifrid, and Aspuru-Guzik]{Tom2024}
G.~Tom, S.~P. Schmid, S.~G. Baird, Y.~Cao, K.~Darvish, H.~Hao, S.~Lo, S.~Pablo-García, E.~M. Rajaonson, M.~Skreta, N.~Yoshikawa, S.~Corapi, G.~D. Akkoc, F.~Strieth-Kalthoff, M.~Seifrid and A.~Aspuru-Guzik, \emph{Chemical Reviews}, 2024, \textbf{124}, 9633--9732\relax
\mciteBstWouldAddEndPuncttrue
\mciteSetBstMidEndSepPunct{\mcitedefaultmidpunct}
{\mcitedefaultendpunct}{\mcitedefaultseppunct}\relax
\EndOfBibitem
\bibitem[Datta and Grant(2004)]{datta2004crystal}
S.~Datta and D.~J. Grant, \emph{Nature Reviews Drug Discovery}, 2004, \textbf{3}, 42--57\relax
\mciteBstWouldAddEndPuncttrue
\mciteSetBstMidEndSepPunct{\mcitedefaultmidpunct}
{\mcitedefaultendpunct}{\mcitedefaultseppunct}\relax
\EndOfBibitem
\bibitem[Beran(2023)]{beran2023frontiers}
G.~J. Beran, \emph{Chemical Science}, 2023, \textbf{14}, 13290--13312\relax
\mciteBstWouldAddEndPuncttrue
\mciteSetBstMidEndSepPunct{\mcitedefaultmidpunct}
{\mcitedefaultendpunct}{\mcitedefaultseppunct}\relax
\EndOfBibitem
\bibitem[Karamertzanis and Pantelides(2007)]{karamertzanis2007ab}
P.~Karamertzanis and C.~Pantelides, \emph{Molecular Physics}, 2007, \textbf{105}, 273--291\relax
\mciteBstWouldAddEndPuncttrue
\mciteSetBstMidEndSepPunct{\mcitedefaultmidpunct}
{\mcitedefaultendpunct}{\mcitedefaultseppunct}\relax
\EndOfBibitem
\bibitem[Reilly \emph{et~al.}(2016)Reilly, Cooper, Adjiman, Bhattacharya, Boese, Brandenburg, Bygrave, Bylsma, Campbell, Car,\emph{et~al.}]{reilly2016report}
A.~M. Reilly, R.~I. Cooper, C.~S. Adjiman, S.~Bhattacharya, A.~D. Boese, J.~G. Brandenburg, P.~J. Bygrave, R.~Bylsma, J.~E. Campbell, R.~Car \emph{et~al.}, \emph{Acta Crystallographica Section B: Structural Science, Crystal Engineering and Materials}, 2016, \textbf{72}, 439--459\relax
\mciteBstWouldAddEndPuncttrue
\mciteSetBstMidEndSepPunct{\mcitedefaultmidpunct}
{\mcitedefaultendpunct}{\mcitedefaultseppunct}\relax
\EndOfBibitem
\bibitem[Neumann \emph{et~al.}(2008)Neumann, Leusen, and Kendrick]{neumann2008major}
M.~A. Neumann, F.~J. Leusen and J.~Kendrick, \emph{Angewandte Chemie International Edition}, 2008, \textbf{47}, 2427--2430\relax
\mciteBstWouldAddEndPuncttrue
\mciteSetBstMidEndSepPunct{\mcitedefaultmidpunct}
{\mcitedefaultendpunct}{\mcitedefaultseppunct}\relax
\EndOfBibitem
\bibitem[Marom \emph{et~al.}(2013)Marom, DiStasio~Jr, Atalla, Levchenko, Reilly, Chelikowsky, Leiserowitz, and Tkatchenko]{marom2013many}
N.~Marom, R.~A. DiStasio~Jr, V.~Atalla, S.~Levchenko, A.~M. Reilly, J.~R. Chelikowsky, L.~Leiserowitz and A.~Tkatchenko, \emph{Angewandte Chemie International Edition}, 2013, \textbf{52}, 6629--6632\relax
\mciteBstWouldAddEndPuncttrue
\mciteSetBstMidEndSepPunct{\mcitedefaultmidpunct}
{\mcitedefaultendpunct}{\mcitedefaultseppunct}\relax
\EndOfBibitem
\bibitem[Firaha \emph{et~al.}(2023)Firaha, Liu, van~de Streek, Sasikumar, Dietrich, Helfferich, Aerts, Braun, Broo, DiPasquale, Lee, Le~Meur, Nilsson~Lill, Lunsmann, Mattei, Muglia, Putra, Raoui, Reutzel-Edens, Rome, Sheikh, Tkatchenko, Woollam, and Neumann]{Firaha2023}
D.~Firaha, Y.~M. Liu, J.~van~de Streek, K.~Sasikumar, H.~Dietrich, J.~Helfferich, L.~Aerts, D.~E. Braun, A.~Broo, A.~G. DiPasquale, A.~Y. Lee, S.~Le~Meur, S.~O. Nilsson~Lill, W.~J. Lunsmann, A.~Mattei, P.~Muglia, O.~D. Putra, M.~Raoui, S.~M. Reutzel-Edens, S.~Rome, A.~Y. Sheikh, A.~Tkatchenko, G.~R. Woollam and M.~A. Neumann, \emph{Nature}, 2023, \textbf{623}, 324--328\relax
\mciteBstWouldAddEndPuncttrue
\mciteSetBstMidEndSepPunct{\mcitedefaultmidpunct}
{\mcitedefaultendpunct}{\mcitedefaultseppunct}\relax
\EndOfBibitem
\bibitem[Musil \emph{et~al.}(2018)Musil, De, Yang, Campbell, Day, and Ceriotti]{musil2018machine}
F.~Musil, S.~De, J.~Yang, J.~E. Campbell, G.~M. Day and M.~Ceriotti, \emph{Chemical science}, 2018, \textbf{9}, 1289--1300\relax
\mciteBstWouldAddEndPuncttrue
\mciteSetBstMidEndSepPunct{\mcitedefaultmidpunct}
{\mcitedefaultendpunct}{\mcitedefaultseppunct}\relax
\EndOfBibitem
\bibitem[Butler \emph{et~al.}(2024)Butler, Hafizi, and Day]{butler2024machine}
P.~W. Butler, R.~Hafizi and G.~M. Day, \emph{The Journal of Physical Chemistry A}, 2024, \textbf{128}, 945--957\relax
\mciteBstWouldAddEndPuncttrue
\mciteSetBstMidEndSepPunct{\mcitedefaultmidpunct}
{\mcitedefaultendpunct}{\mcitedefaultseppunct}\relax
\EndOfBibitem
\bibitem[Podryabinkin \emph{et~al.}(2019)Podryabinkin, Tikhonov, Shapeev, and Oganov]{PhysRevB.99.064114}
E.~V. Podryabinkin, E.~V. Tikhonov, A.~V. Shapeev and A.~R. Oganov, \emph{Phys. Rev. B}, 2019, \textbf{99}, 064114\relax
\mciteBstWouldAddEndPuncttrue
\mciteSetBstMidEndSepPunct{\mcitedefaultmidpunct}
{\mcitedefaultendpunct}{\mcitedefaultseppunct}\relax
\EndOfBibitem
\bibitem[Gubaev \emph{et~al.}(2019)Gubaev, Podryabinkin, Hart, and Shapeev]{Gubaev2019}
K.~Gubaev, E.~V. Podryabinkin, G.~L. Hart and A.~V. Shapeev, \emph{Computational Materials Science}, 2019, \textbf{156}, 148--156\relax
\mciteBstWouldAddEndPuncttrue
\mciteSetBstMidEndSepPunct{\mcitedefaultmidpunct}
{\mcitedefaultendpunct}{\mcitedefaultseppunct}\relax
\EndOfBibitem
\bibitem[Wang \emph{et~al.}(2022)Wang, Liu, Lile, Norwood, Hernandez, Manna, and Mueller]{Wang2022}
Y.~Wang, S.~Liu, P.~Lile, S.~Norwood, A.~Hernandez, S.~Manna and T.~Mueller, \emph{npj Computational Materials}, 2022, \textbf{8}, 173\relax
\mciteBstWouldAddEndPuncttrue
\mciteSetBstMidEndSepPunct{\mcitedefaultmidpunct}
{\mcitedefaultendpunct}{\mcitedefaultseppunct}\relax
\EndOfBibitem
\bibitem[Manna \emph{et~al.}(2023)Manna, Wang, Hernandez, Lile, Liu, and Mueller]{Manna2023}
S.~Manna, Y.~Wang, A.~Hernandez, P.~Lile, S.~Liu and T.~Mueller, \emph{Scientific Data}, 2023, \textbf{10}, 308\relax
\mciteBstWouldAddEndPuncttrue
\mciteSetBstMidEndSepPunct{\mcitedefaultmidpunct}
{\mcitedefaultendpunct}{\mcitedefaultseppunct}\relax
\EndOfBibitem
\bibitem[Shapeev(2016)]{Shapeev2016}
A.~V. Shapeev, \emph{Multiscale Modeling \& Simulation}, 2016, \textbf{14}, 1153--1173\relax
\mciteBstWouldAddEndPuncttrue
\mciteSetBstMidEndSepPunct{\mcitedefaultmidpunct}
{\mcitedefaultendpunct}{\mcitedefaultseppunct}\relax
\EndOfBibitem
\bibitem[Podryabinkin and Shapeev(2017)]{Podryabinkin2017}
E.~V. Podryabinkin and A.~V. Shapeev, \emph{Computational Materials Science}, 2017, \textbf{140}, 171--180\relax
\mciteBstWouldAddEndPuncttrue
\mciteSetBstMidEndSepPunct{\mcitedefaultmidpunct}
{\mcitedefaultendpunct}{\mcitedefaultseppunct}\relax
\EndOfBibitem
\bibitem[Kitaigorodskii(2003)]{kitaigorodskii2003organic}
A.~Kitaigorodskii, \emph{J. Appl. Crystallogr}, 2003, \textbf{36}, 7--13\relax
\mciteBstWouldAddEndPuncttrue
\mciteSetBstMidEndSepPunct{\mcitedefaultmidpunct}
{\mcitedefaultendpunct}{\mcitedefaultseppunct}\relax
\EndOfBibitem
\bibitem[Zhu and Hattori(2023)]{zhu2023organic}
Q.~Zhu and S.~Hattori, \emph{Journal of Materials Research}, 2023, \textbf{38}, 19--36\relax
\mciteBstWouldAddEndPuncttrue
\mciteSetBstMidEndSepPunct{\mcitedefaultmidpunct}
{\mcitedefaultendpunct}{\mcitedefaultseppunct}\relax
\EndOfBibitem
\bibitem[Fredericks \emph{et~al.}(2021)Fredericks, Parrish, Sayre, and Zhu]{fredericks2021pyxtal}
S.~Fredericks, K.~Parrish, D.~Sayre and Q.~Zhu, \emph{Computer Physics Communications}, 2021, \textbf{261}, 107810\relax
\mciteBstWouldAddEndPuncttrue
\mciteSetBstMidEndSepPunct{\mcitedefaultmidpunct}
{\mcitedefaultendpunct}{\mcitedefaultseppunct}\relax
\EndOfBibitem
\bibitem[Rybin and Shapeev(2024)]{rybin2024moment}
N.~Rybin and A.~Shapeev, \emph{Journal of Applied Physics}, 2024, \textbf{135}, 205108\relax
\mciteBstWouldAddEndPuncttrue
\mciteSetBstMidEndSepPunct{\mcitedefaultmidpunct}
{\mcitedefaultendpunct}{\mcitedefaultseppunct}\relax
\EndOfBibitem
\bibitem[Rybin \emph{et~al.}(2024)Rybin, Maksimov, Zaikov, and Shapeev]{rybin2024thermophysical}
N.~Rybin, D.~Maksimov, Y.~Zaikov and A.~Shapeev, \emph{Journal of Molecular Liquids}, 2024, \textbf{410}, 125402\relax
\mciteBstWouldAddEndPuncttrue
\mciteSetBstMidEndSepPunct{\mcitedefaultmidpunct}
{\mcitedefaultendpunct}{\mcitedefaultseppunct}\relax
\EndOfBibitem
\bibitem[Adamo and Barone(1999)]{Adamo1999}
C.~Adamo and V.~Barone, \emph{The Journal of Chemical Physics}, 1999, \textbf{110}, 6158--6170\relax
\mciteBstWouldAddEndPuncttrue
\mciteSetBstMidEndSepPunct{\mcitedefaultmidpunct}
{\mcitedefaultendpunct}{\mcitedefaultseppunct}\relax
\EndOfBibitem
\bibitem[Tkatchenko \emph{et~al.}(2012)Tkatchenko, DiStasio, Car, and Scheffler]{PhysRevLett.108.236402}
A.~Tkatchenko, R.~A. DiStasio, R.~Car and M.~Scheffler, \emph{Phys. Rev. Lett.}, 2012, \textbf{108}, 236402\relax
\mciteBstWouldAddEndPuncttrue
\mciteSetBstMidEndSepPunct{\mcitedefaultmidpunct}
{\mcitedefaultendpunct}{\mcitedefaultseppunct}\relax
\EndOfBibitem
\bibitem[Yang \emph{et~al.}(2014)Yang, Hu, Usvyat, Matthews, Schütz, and Chan]{Yang2014}
J.~Yang, W.~Hu, D.~Usvyat, D.~Matthews, M.~Schütz and G.~K.-L. Chan, \emph{Science}, 2014, \textbf{345}, 640--643\relax
\mciteBstWouldAddEndPuncttrue
\mciteSetBstMidEndSepPunct{\mcitedefaultmidpunct}
{\mcitedefaultendpunct}{\mcitedefaultseppunct}\relax
\EndOfBibitem
\bibitem[Novikov \emph{et~al.}(2020)Novikov, Gubaev, Podryabinkin, and Shapeev]{Novikov2021}
I.~S. Novikov, K.~Gubaev, E.~V. Podryabinkin and A.~V. Shapeev, \emph{Machine Learning: Science and Technology}, 2020, \textbf{2}, 025002\relax
\mciteBstWouldAddEndPuncttrue
\mciteSetBstMidEndSepPunct{\mcitedefaultmidpunct}
{\mcitedefaultendpunct}{\mcitedefaultseppunct}\relax
\EndOfBibitem
\bibitem[Goreinov \emph{et~al.}(2010)Goreinov, Oseledets, Savostyanov, Tyrtyshnikov, and Zamarashkin]{zamarashkin2010-maxvol}
S.~A. Goreinov, I.~V. Oseledets, D.~V. Savostyanov, E.~E. Tyrtyshnikov and N.~L. Zamarashkin, \emph{Matrix Methods: Theory, Algorithms And Applications: Dedicated to the Memory of Gene Golub}, World Scientific, 2010, pp. 247--256\relax
\mciteBstWouldAddEndPuncttrue
\mciteSetBstMidEndSepPunct{\mcitedefaultmidpunct}
{\mcitedefaultendpunct}{\mcitedefaultseppunct}\relax
\EndOfBibitem
\bibitem[Kresse and Furthm\"uller(1996)]{Kresse1996}
G.~Kresse and J.~Furthm\"uller, \emph{Phys. Rev. B}, 1996, \textbf{54}, 11169--11186\relax
\mciteBstWouldAddEndPuncttrue
\mciteSetBstMidEndSepPunct{\mcitedefaultmidpunct}
{\mcitedefaultendpunct}{\mcitedefaultseppunct}\relax
\EndOfBibitem
\bibitem[Kresse and Joubert(1999)]{Kresse1999}
G.~Kresse and D.~Joubert, \emph{Phys. Rev. B}, 1999, \textbf{59}, 1758--1775\relax
\mciteBstWouldAddEndPuncttrue
\mciteSetBstMidEndSepPunct{\mcitedefaultmidpunct}
{\mcitedefaultendpunct}{\mcitedefaultseppunct}\relax
\EndOfBibitem
\bibitem[Perdew \emph{et~al.}(1996)Perdew, Burke, and Ernzerhof]{Perdew1996}
J.~P. Perdew, K.~Burke and M.~Ernzerhof, \emph{Phys. Rev. Lett.}, 1996, \textbf{77}, 3865--3868\relax
\mciteBstWouldAddEndPuncttrue
\mciteSetBstMidEndSepPunct{\mcitedefaultmidpunct}
{\mcitedefaultendpunct}{\mcitedefaultseppunct}\relax
\EndOfBibitem
\bibitem[Grimme \emph{et~al.}(2010)Grimme, Antony, Ehrlich, and Krieg]{Grimme2010}
S.~Grimme, J.~Antony, S.~Ehrlich and H.~Krieg, \emph{The Journal of Chemical Physics}, 2010, \textbf{132}, 154104\relax
\mciteBstWouldAddEndPuncttrue
\mciteSetBstMidEndSepPunct{\mcitedefaultmidpunct}
{\mcitedefaultendpunct}{\mcitedefaultseppunct}\relax
\EndOfBibitem
\bibitem[Ciabini \emph{et~al.}(2005)Ciabini, Gorelli, Santoro, Bini, Schettino, and Mezouar]{ciabini2005high}
L.~Ciabini, F.~A. Gorelli, M.~Santoro, R.~Bini, V.~Schettino and M.~Mezouar, \emph{Physical Review B—Condensed Matter and Materials Physics}, 2005, \textbf{72}, 094108\relax
\mciteBstWouldAddEndPuncttrue
\mciteSetBstMidEndSepPunct{\mcitedefaultmidpunct}
{\mcitedefaultendpunct}{\mcitedefaultseppunct}\relax
\EndOfBibitem
\bibitem[Ciabini \emph{et~al.}(2007)Ciabini, Santoro, Gorelli, Bini, Schettino, and Raugei]{ciabini2007triggering}
L.~Ciabini, M.~Santoro, F.~A. Gorelli, R.~Bini, V.~Schettino and S.~Raugei, \emph{Nature materials}, 2007, \textbf{6}, 39--43\relax
\mciteBstWouldAddEndPuncttrue
\mciteSetBstMidEndSepPunct{\mcitedefaultmidpunct}
{\mcitedefaultendpunct}{\mcitedefaultseppunct}\relax
\EndOfBibitem
\bibitem[Wen \emph{et~al.}(2011)Wen, Hoffmann, and Ashcroft]{wen2011benzene}
X.-D. Wen, R.~Hoffmann and N.~Ashcroft, \emph{Journal of the American Chemical Society}, 2011, \textbf{133}, 9023--9035\relax
\mciteBstWouldAddEndPuncttrue
\mciteSetBstMidEndSepPunct{\mcitedefaultmidpunct}
{\mcitedefaultendpunct}{\mcitedefaultseppunct}\relax
\EndOfBibitem
\bibitem[Raiteri \emph{et~al.}(2005)Raiteri, Marto{\v{n}}{\'a}k, and Parrinello]{raiteri2005exploring}
P.~Raiteri, R.~Marto{\v{n}}{\'a}k and M.~Parrinello, \emph{Angewandte Chemie International Edition}, 2005, \textbf{44}, 3769--3773\relax
\mciteBstWouldAddEndPuncttrue
\mciteSetBstMidEndSepPunct{\mcitedefaultmidpunct}
{\mcitedefaultendpunct}{\mcitedefaultseppunct}\relax
\EndOfBibitem
\bibitem[Finkelstein and Ptitsyn(2016)]{finkelstein2016protein}
A.~V. Finkelstein and O.~Ptitsyn, \emph{Protein physics: a course of lectures}, Elsevier, 2016\relax
\mciteBstWouldAddEndPuncttrue
\mciteSetBstMidEndSepPunct{\mcitedefaultmidpunct}
{\mcitedefaultendpunct}{\mcitedefaultseppunct}\relax
\EndOfBibitem
\bibitem[Zhu \emph{et~al.}(2012)Zhu, Oganov, Glass, and Stokes]{zhu2012constrained}
Q.~Zhu, A.~R. Oganov, C.~W. Glass and H.~T. Stokes, \emph{Acta Crystallographica Section B: Structural Science}, 2012, \textbf{68}, 215--226\relax
\mciteBstWouldAddEndPuncttrue
\mciteSetBstMidEndSepPunct{\mcitedefaultmidpunct}
{\mcitedefaultendpunct}{\mcitedefaultseppunct}\relax
\EndOfBibitem
\bibitem[Chisholm \emph{et~al.}(2005)Chisholm, Motherwell, Tulip, Parsons, and Clark]{chisholm2005ab}
J.~A. Chisholm, S.~Motherwell, P.~R. Tulip, S.~Parsons and S.~J. Clark, \emph{Crystal growth \& design}, 2005, \textbf{5}, 1437--1442\relax
\mciteBstWouldAddEndPuncttrue
\mciteSetBstMidEndSepPunct{\mcitedefaultmidpunct}
{\mcitedefaultendpunct}{\mcitedefaultseppunct}\relax
\EndOfBibitem
\bibitem[Hamad \emph{et~al.}(2008)Hamad, Hughes, Catlow, and Harris]{hamad2008clustering}
S.~Hamad, C.~E. Hughes, C.~R.~A. Catlow and K.~D. Harris, \emph{The Journal of Physical Chemistry B}, 2008, \textbf{112}, 7280--7288\relax
\mciteBstWouldAddEndPuncttrue
\mciteSetBstMidEndSepPunct{\mcitedefaultmidpunct}
{\mcitedefaultendpunct}{\mcitedefaultseppunct}\relax
\EndOfBibitem
\bibitem[Dawson \emph{et~al.}(2005)Dawson, Allan, Belmonte, Clark, David, McGregor, Parsons, Pulham, and Sawyer]{dawson2005effect}
A.~Dawson, D.~R. Allan, S.~A. Belmonte, S.~J. Clark, W.~I. David, P.~A. McGregor, S.~Parsons, C.~R. Pulham and L.~Sawyer, \emph{Crystal Growth and Design}, 2005, \textbf{5}, 1415--1427\relax
\mciteBstWouldAddEndPuncttrue
\mciteSetBstMidEndSepPunct{\mcitedefaultmidpunct}
{\mcitedefaultendpunct}{\mcitedefaultseppunct}\relax
\EndOfBibitem
\bibitem[Moggach \emph{et~al.}(2008)Moggach, Parsons, and Wood]{moggach2008high}
S.~A. Moggach, S.~Parsons and P.~A. Wood, \emph{Crystallography reviews}, 2008, \textbf{14}, 143--184\relax
\mciteBstWouldAddEndPuncttrue
\mciteSetBstMidEndSepPunct{\mcitedefaultmidpunct}
{\mcitedefaultendpunct}{\mcitedefaultseppunct}\relax
\EndOfBibitem
\bibitem[Boldyreva \emph{et~al.}(2003)Boldyreva, Drebushchak, and Shutova]{boldyreva2003structural}
E.~Boldyreva, T.~Drebushchak and E.~Shutova, \emph{Zeitschrift f{\"u}r Kristallographie-Crystalline Materials}, 2003, \textbf{218}, 366--376\relax
\mciteBstWouldAddEndPuncttrue
\mciteSetBstMidEndSepPunct{\mcitedefaultmidpunct}
{\mcitedefaultendpunct}{\mcitedefaultseppunct}\relax
\EndOfBibitem
\bibitem[Perlovich \emph{et~al.}(2001)Perlovich, Hansen, and Bauer-Brandl]{perlovich2001polymorphism}
G.~Perlovich, L.~K. Hansen and A.~Bauer-Brandl, \emph{Journal of thermal analysis and calorimetry}, 2001, \textbf{66}, 699--715\relax
\mciteBstWouldAddEndPuncttrue
\mciteSetBstMidEndSepPunct{\mcitedefaultmidpunct}
{\mcitedefaultendpunct}{\mcitedefaultseppunct}\relax
\EndOfBibitem
\bibitem[Thiéry and Léger(1988)]{10.1063/1.454809}
M.~M. Thiéry and J.~M. Léger, \emph{The Journal of Chemical Physics}, 1988, \textbf{89}, 4255--4271\relax
\mciteBstWouldAddEndPuncttrue
\mciteSetBstMidEndSepPunct{\mcitedefaultmidpunct}
{\mcitedefaultendpunct}{\mcitedefaultseppunct}\relax
\EndOfBibitem
\bibitem[Hodapp and Shapeev(2023)]{hodapp2023equivariant}
M.~Hodapp and A.~Shapeev, \emph{arXiv preprint arXiv:2304.08226}, 2023\relax
\mciteBstWouldAddEndPuncttrue
\mciteSetBstMidEndSepPunct{\mcitedefaultmidpunct}
{\mcitedefaultendpunct}{\mcitedefaultseppunct}\relax
\EndOfBibitem
\bibitem[Ko \emph{et~al.}(2021)Ko, Finkler, Goedecker, and Behler]{Ko2021}
T.~W. Ko, J.~A. Finkler, S.~Goedecker and J.~Behler, \emph{Nature Communications}, 2021, \textbf{12}, 398\relax
\mciteBstWouldAddEndPuncttrue
\mciteSetBstMidEndSepPunct{\mcitedefaultmidpunct}
{\mcitedefaultendpunct}{\mcitedefaultseppunct}\relax
\EndOfBibitem
\bibitem[Shaidu \emph{et~al.}(2024)Shaidu, Pellegrini, K{\"u}{\c{c}}{\"u}kbenli, Lot, and de~Gironcoli]{Shaidu2024}
Y.~Shaidu, F.~Pellegrini, E.~K{\"u}{\c{c}}{\"u}kbenli, R.~Lot and S.~de~Gironcoli, \emph{npj Computational Materials}, 2024, \textbf{10}, 47\relax
\mciteBstWouldAddEndPuncttrue
\mciteSetBstMidEndSepPunct{\mcitedefaultmidpunct}
{\mcitedefaultendpunct}{\mcitedefaultseppunct}\relax
\EndOfBibitem
\bibitem[Grisafi and Ceriotti(2019)]{Grisafi2019}
A.~Grisafi and M.~Ceriotti, \emph{The Journal of Chemical Physics}, 2019, \textbf{151}, 204105\relax
\mciteBstWouldAddEndPuncttrue
\mciteSetBstMidEndSepPunct{\mcitedefaultmidpunct}
{\mcitedefaultendpunct}{\mcitedefaultseppunct}\relax
\EndOfBibitem
\bibitem[Kapil and Engel(2022)]{kapil2022complete}
V.~Kapil and E.~A. Engel, \emph{Proceedings of the National Academy of Sciences}, 2022, \textbf{119}, e2111769119\relax
\mciteBstWouldAddEndPuncttrue
\mciteSetBstMidEndSepPunct{\mcitedefaultmidpunct}
{\mcitedefaultendpunct}{\mcitedefaultseppunct}\relax
\EndOfBibitem
\bibitem[Novikov \emph{et~al.}(2024)Novikov, Makarov, Suleimanov, and Shapeev]{novikov2024-rpmd-oh-hbr}
I.~S. Novikov, E.~M. Makarov, Y.~V. Suleimanov and A.~V. Shapeev, \emph{Chemical Physics Letters}, 2024,  141620\relax
\mciteBstWouldAddEndPuncttrue
\mciteSetBstMidEndSepPunct{\mcitedefaultmidpunct}
{\mcitedefaultendpunct}{\mcitedefaultseppunct}\relax
\EndOfBibitem
\end{mcitethebibliography}
\bibliographystyle{rsc} 
\end{document}